\documentclass[onecollarge,runningheads]{svjour2}
\smartqed  
\usepackage{graphicx}
\journalname{Astrophysics and Space Science}
\begin{document}
\title{Expulsion   of   Magnetic  Flux  Lines  from  the Growing
Superconducting Core of a Magnetised Quark Star}
\titlerunning{Expulsion of magnetic lines of force from quark star}
\author{Somenath Chakrabarty}
\institute{
Department of Physics, Visva-Bharati\\ Santiniketan 731 235,\\ 
West Bengal, India\\ E-Mail:somenath.chakrabarty@visva-bharati.ac.in}
\date{Received: date / Accepted: date}

\maketitle
\begin{abstract}
The  expulsion  of  magnetic flux lines from a growing
superconducting core of a quark star has been investigated.
The idea  of impurity diffusion in molten alloys and an identical mechanism
of baryon number transport from hot quark-gluon-plasma phase to
hadronic phase during quark-hadron phase transition in the early universe,
micro-second after big bang has been used. The possibility
of Mullins-Sekerka normal-superconducting  interface  instability
has also been studied.
\end{abstract}

\section{Introduction}
If the matter density at the core of a neutron star exceeds a few
times    normal    nuclear    density    (e.g.   $>3n_0$,   where
$n_0=0.17$fm$^{-3}$, the normal nuclear density),  a  deconfining
phase transition to quark matter may occur. As a consequence
a  normal neutron star will be converted to a hybrid star with an
infinite cluster of quark matter core  and  a  crust  of  neutron
matter. If the speculation of Witten \cite{R1} that a flavour symmetric
quark matter may be the absolute ground state at zero temperature
and  pressure is assumed to be correct, then it may be quite possible that 
the whole neutron star / hybrid star will be converted to a star of quark matter, 
known as quark star or strange  star.

From  the observed features in the spectra of pulsating accreting
neutron stars in binary system, the strength of surface  magnetic
field  of a normal neutron star is found to be $\sim 10^{12}$G
\cite{R2,R3,R4,R5}. At the core region of a neutron  star  it  
probably  reaches a few orders of magnitude more than the surface value.  
In some publications several years ago we have shown that if the intensity of 
magnetic field at the core region of a compact neutron star exceeds some 
critical value which is the typical strength of magnetic field at which the
cyclotron lines begin to  appear  or  equivalently  at  which  the
cyclotron  quantum  is  of  the order of or greater than the rest
mass of the particle considered or the de Broglie wave length  is
of  the  order  of  or  greater  than  the  Larmor  radius of the
particle, then there can not be nucleation of any   quark  matter
bubble  in  the metastable neutron matter \cite{R6,R7}. The surface as well as
the curvature energies of the new phase diverge in this case. As a consequence the
new quark matter phase  can  not  be  thermodynamically
favourable  over the metastable neutron matter phase. Therefore to
achieve a first order deconfining  transition  initiated  by  the
nucleation  of  quark  droplets  at  the core of neutron star, it has to
be assumed that the strength of magnetic field throughout the star is
much less than the corresponding critical value. In the  case  of
electron of mass $0.5$MeV, this critical field is $\sim 4.4\times
10^{13}$G,  for  light quarks of current mass $5$MeV, it is $\sim
10^{15}$G, whereas for $s$-quark of current mass $150$MeV, it  is
$\sim 10^{20}$G.

Now  for  a  many  body fermion system, the microscopic theory of
superconductivity suggests that if  the  interaction  favours
formation  of  pairs at low temperature \cite{R8}, the system may undergo a
phase transition to a super-fluid state. This is expected to occur
in the dense neutron matter present in neutron star \cite{R9,R10,R11}. On the
other hand, if the particles carry charges, the paired state  will
be  superconducting (neutron matter becomes super-fluid, whereas the small
percentage of protons undergoes a transition to type-II superconducting
phase).  In  the  case  of  a many body system of electrons
the well known BCS theory is generally used to study the superconducting 
properties due to electron pairing \cite{R8}.  One
electron  of  momentum  $\vec  k$ and spin $\vec s$ combines with
another one of momentum $-\vec k$ and spin $-\vec s$ and  form  a
Cooper  pair.  The  coupling  is  mediated by the electron-phonon
interaction.  In  the  case  of  quark matter,  the  basic
quark-quark  interaction  is  attractive  at  large  distances and
consequently the BCS pairing mechanism is also  applicable  here.
For  a  highly  degenerate system, which is true in quark star,
the pairing  takes  place  near  the  Fermi  surface.  The  other
condition  that must be satisfied to form Cooper pairs is that the
temperature ($T$) of the system should  be  much  less  than  the
superconducting energy gap ($\Delta$), which is a function of the
interaction  strength  and the density of the system. This is the
most important criterion for  the  occurrence  of  superconducting
transition.  In  the  case  of quark stars, only quarks can form
Cooper pairs. Whereas the electrons, whose density is extremely low
compared to quark matter, may be treated as highly degenerate  relativistic  
plasma and are  unlikely  to  form Cooper pairs (since the quark matter is in $\beta$-equilibrium, a small fraction of electrons should be present in the system). The kinetic energy of the 
electronic part dominates over its attractive potential energy (which is 
electromagnetic in nature), and  as  a  result the
corresponding superconducting transition temperature should be extremely
low and in reality may not be achieved in a quark star. The relativistic
version of the theory of super-fluidity and superconductivity for  a 
system of many fermions  was developed long ago by Bailin and Love \cite{R12}.
In that paper, an overview of mathematical formalism for 
relativistic version of BCS theory is given and discussed with elaborate 
mathematical derivation to obtain the critical field(s) for both type-I and 
type-II superconductors. They have also given a rigorous mathematical derivation 
to get the critical temperature above which the superconducting property is 
completely destroyed and also obtained the expressions for correlation
lengths.

In the past few years a lot of works have also been done on a new concept, the
possibility of colour superconductivity in quark stars with light flavor
pairing, known as $2SC$, with unpaired $s$-quarks. A lot of works have also
been done on the possibility of an entirely new kind of phase, known as the
{\it{Colour-Flavor Locked}} (CFL) phase, which is colour neutral and 
charge neutral system and is also expected to be energetically more favourable than 
strange quark matter \cite{R13,R14,R15,R16,R17,R18,R19,R19a}.
However, the density at which this new phase appears is several times normal
nuclear density.

In the present article we have assumed a type-I superconducting phase transition in
quark matter at the core region of a quark star and investigated the mechanism by 
which the magnetic flux lines are expelled from the superconducting zone. 

In a very recent work by Konenkov and Geppert have investigated the
expulsion of magnetic flux lines from superconducting core region of
neutron stars \cite{R20}. They have considered a type-II
superconducting transition at the core region and studied the movement
of quantized fluxoids. They have also given a mechanism by which 
the flux lines expelled from the core into the crustal region undergo ohmic decay.

Now  the  quarks with identical Fermi energy can only combine to form
Cooper pairs at the Fermi surface. Since for $u$ and $d$ quarks, the current masses are 
equal and also their chemical potentials are almost identical, whereas $s$
quark is much more heavier than  $u$  and  $d$  quarks  and  also  its
chemical  potential  is  different,  we may therefore  have only $uu$, $dd$,
$ud$ and $ss$ Cooper pairs in  the  system.  For  iso-spin  $1/2$
flavours,  the  contribution  may  come either from iso-scalar or
iso-vector channels. It was shown in ref.\cite{R12} that  the  pairing
of $u-u$, $d-d$ or $u-d$  system  will  be favoured by iso-scalar combination
rather than iso-vector channel.  On  the  other  hand  the  $s-s$
combination  is  a  triplet  state  with  $J^P=1^+$. It has been shown in
ref.\cite{R12} that if a normal quark matter  system
undergoes  a superconducting phase transition, the newly produced
quark matter phase will be a type-I superconductor. They have also shown in 
that 
paper that the critical magnetic field to destroy such pairing is  $\sim  10^{16}$G
for $n\sim 2-3n_0$, which is indeed much larger than the typical
pulsar magnetic field 
The corresponding critical temperature  is
$\sim 10^9-10^{10}$K, which is again high enough for quark stars / core
of hybrid stars, which are expected to be extremely cold objects. 
(in this connection we would like to mention
that in the present article we are not going to study the
superconducting properties of the matter inside the strange stellar objects, 
called the magnetars, with the surface magnetic field strength $\geq
10^{15}$G). 
Therefore, we may expect that the magnetic field strengths at the core
region of a quark star are much less than the corresponding critical
value for the destruction of superconducting property and the temperature
is also low enough. 
Then during such a type-I superconducting phase
transition, the  magnetic  flux  lines  from the superconducting
quark sector of the quark star will be pushed out  towards  the
normal  crustal  region.  Now for a small type-I superconducting
laboratory sample placed in an external magnetic field less than the
corresponding critical
value, the expulsion  of  magnetic  field  takes  place
instantaneously. Whereas in the quark star scenario, the picture may 
be completely different. It may take several thousands of years for the 
magnetic flux
lines to get expelled from the superconducting core region. Which
further means, that the growth of superconducting phase in quark stars
will not be instantaneous.
Therefore in a quark star / hybrid star, with type-I superconducting
quark matter at the core, the magnetic flux lines will be completely
expelled by Meissner effect not instantaneously, it takes several thousand
years of time. A simple
estimate shows that the expulsion time due to ohmic diffusion is $\sim
10^4$yrs \cite{R22}. It was shown by Chau using Ginzberg-Landau
formalism that the time for expulsion of magnetic lines of force
accompanied by the enhancement of magnetic 
field non-uniformities at the crustal region gets prolonged to $10^7$yrs
\cite{R23}. Alford et. al. investigated the expulsion of magnetic lines
of force from the colour superconducting region by considering the pairing of
like and unlike quarks and obtained the expulsion time much larger than
the age of the Universe \cite{R24}.

The  aim  of the present article is to investigate the expulsion of magnetic
flux lines from the growing superconducting core of a quark star.
We have used the idea of impurity diffusion in molten alloys or the transport of 
baryon numbers from hot quark matter soup to hadronic matter during quark-hadron 
phase transition in the early universe, micro-second after big bang (the first
mechanism is used by the material scientists and metallurgists \cite{R24a}, 
whereas the
later one is used by cosmologists working in the field of big bang
nucleo-synthesis \cite{R25,R25a}), We  have also studied the possibility of 
Mullins-Sekerka normal-superconducting interface instability \cite{R26a,R26b} 
in quark
matter. This is generally observed in the case of solidification of pure molten 
metals at the solid-liquid interface, if there is a temperature gradient. 
The interface will always be stable if the  temperature  gradient  is  positive  
and otherwise it will be unstable. In alloys, the criteria for stable / unstable 
behaviour is more complicated. It is seen that, during the solidification of an
alloy, there is a substantial change in the  concentration  ahead
of the interface. Here solute diffusion as well as the heat flow
effects must be considered simultaneously. The
particular  problem  we are going to investigate here is analogous to
solute diffusion during solidification of an alloy.

\section{Formalism}
It  has  been  assumed  that  the growth of superconducting quark
bubble started from the centre of the star  and  the
nomenclature {\it{controlled growth}} for such phenomenon has been used. If the
magnetic  field strength and the temperature of the star are a few
orders of magnitude 
less  than  their  critical  values,  the  normal  quark matter  phase  is
thermodynamically   unstable   relative   to   the  corresponding
superconducting one.  Then  due  to  fluctuation,  a  droplet  of
superconducting quark matter bubble may be nucleated in metastable
normal quark matter medium.  If  the  size  of this superconducting bubble is 
greater than the  corresponding  critical  value,  it  will  act  as  the
nucleating  centre  for the growth of superconducting quark core.
The critical radius can be obtained by minimising the free energy.
Then following the work of Mullins and Sekerka, we have \cite{R26a,R26b}
\begin{equation}
r_c=\frac{16\pi\alpha}{B_m^{(c)2}\left [1-\left (\frac{B_m}{B_m^{(c)}}\right )^2
\right ]},
\end{equation}
where $\alpha$ is  the  surface tension or the surface energy per unit area of the
critical superconducting bubbles, (from this expression it is possible to 
obtain the critical size of the quark matter bubble by considering $10^{-3}
\leq \alpha 1$ as the range of surface tension) which is greater than zero 
for a type-I
superconductor-normal  interface,  $B_m^{(c)}$  is  the  critical
magnetic field. In presence of a magnetic field $B_m< B_m^{(c)}$,
the  normal  to  superconducting  transition  is  first  order in
nature. As the  superconducting  phase  grows  continuously,  the
magnetic  field  lines  will  be pushed out into the normal quark
matter crust. This is the usual {\it {Meissner effect}}, observed
in type-I superconductor. We compare this phenomenon of  magnetic
flux  expulsion  from a growing superconducting quark matter core with the
diffusion of impurities from the frozen phase of molten metal or the transport 
of baryon numbers from hot quark matter soup during quark-hadron phase
transition in the early universe. The  formation  of
superconducting  zone  is  compared  with  the  solidification of
molten metal or with the transition to hadronic phase with almost zero baryon 
number. It is known from the simple thermodynamic  calculations
that  if the free energy of molten phase decreases in presence of
impurity atoms, then during solidification they prefer to  reside
in  the  molten phase otherwise they go to the solid phase. In this particular 
case the magnetic field lines play the role of impurity atoms and because of 
Meissner effect, they prefer to remain in normal quark matter phase, the normal 
quark  matter phase plays the role of molten metal or the hot quark soup, whereas 
the superconducting phase can be compared with the frozen solid phase or the 
hadronic phase. (This idea was applied to baryon number transport during first order
quark-hadron phase transition in the early Universe, where baryon
number replaces impurity, quark phase replaces molten  metal  and
hadronic  matter replaces that of solid metal \cite{R25,R25a}. The baryon number
prefers to stay in the quark phase because of Boltzmann suppression factor
in the hadronic phase). Since the magnetic flux lines prefer to reside in the 
normal phase, the well known Meissner effect can therefore be restated as 
{\it{the solubility of magnetic flux lines in the superconducting phase is zero  
with a finite penetration depth}}.

The  dynamical  equation  for  the flux expulsion can be obtained
from  the  simplified  model  of   sharp   normal-superconducting
interface.  The  expulsion  equation  is  given by the well known
diffusion equation \cite{R27}
\begin{equation}
\frac{\partial B_m}{\partial t}=D\nabla^2 B_m
\end{equation}
where $B_m$ is the  magnetic  field  intensity  and  $D$  is  the
diffusion coefficient, given by
\begin{equation}
D=\frac{c^2}{4\pi \sigma_n}
\end{equation}
where $\sigma_n$ is the electrical conductivity of the normal quark matter phase 
with $B_m=0$ and the symbol $c$ is the velocity of light.. The electrical conductivity of quark matter for $B_m=0$ is given by 
\cite{R28,R28a}
\[
\sigma_n=5.8\times  10^{25} \left (\frac{\alpha_s}{0.1}\right
)^{-3/2} T_{10}^{-2} \left ( \frac{n}{n_0}\right )
\]
whereas the more appropriate form is \cite{D48}
\begin{equation}
\sigma_n\sim (\alpha_s T)^{-5/3}
\end{equation}
and is expressed in sec$^{-1}$, where $\alpha_s$ is the strong coupling constant
and $T_{10}=T/10^{10}$K, the numerical value for this electrical
conductivity in  the  case  of  quark matter relevant for quark star density  
is  $\sim 10^{26}$ sec$^{-1}$. We have used this expression to get an order
of magnitude estimate of electrical conductivity of quark matter.
In  actual calculation one has to evaluate $\sigma_n$ in presence
of $B_m$. In that case, $\sigma_n$ may not be a scalar quantity 
(magnetic field destroys the isotropy of the electromagnetic properties
of the medium). In particular, for extremely large $B_m$, the components 
of electric current vector orthogonal to $B_m$ become almost zero. Which indicates 
that the quarks can only move along the direction
of magnetic field or in other words, across the field resistivity becomes infinity.
Here, by normal quark matter we mean that it is non-superconducting in
nature.
Further, while calculating the electrical conductivity, we have assumed that
quark matter is in $\beta$-equilibrium. In such condition, since the mass of
$s$-quark is assumed to be $150$Mev, the electron density is not exactly
zero, but it is a few orders of magnitude less than the $s$-quark density.
Therefore, we can neglect the effect of electrical conductivity from
electrons. We have also noticed, that the kinetic energy of these electrons
are very high and as a consequence, we can not have superconducting transition
of electrons in quark stars.

A solution of eqn.(2) with spherical symmetry can be obtained by Greens' function  
technique, and  is  given  by  (for a general topological structure, no
analytical solution is possible)
\begin{eqnarray}
B_m(r,t)&=&\frac{1}{2r (\pi D t)^{1/2}}\int_0^\infty B_m^{(0)}
(r')\nonumber \\ &&\big [ \exp(-u_-^2)-\exp(-u_+^2)\big ]r'dr'
\end{eqnarray}
where  $u_\pm=(r\pm  r')/2(Dt)^{1/2}$  and  $B_m^{(0)}(r)$ is the
magnetic field distribution within the star at $t=0$, which is of
course an entirely unknown function of radial coordinate $r$.  To
obtain an estimate of magnetic field diffusion time scale ($\tau_D$),
we assume $B_m^{(0)}(r)=B_m^{(0)}=$ constant (in reality, this may
not  be true  inside the star, e.g., in some work we took a parametric
form, given by
\[
B_m(n_B)=B_m^{\rm{surf.}}+B_0\left [ 1-\exp\left (-\beta \left (
\frac{n_B}{n_0}\right )^\gamma \right ) \right ]
\]
with $\beta=0.01$, $\gamma=3$, $B_0=5\times 10^{18}$G and
$B_m^{\rm{Surf.}}\approx 10^{14}$. We have used this radial form of distribution
to study the mean field properties of dense quark matter in presence of strong 
quantising magnetic field \cite{R29,R29a}. However, 
such a parametrisation is very difficult to use in this particular case). Then 
using the expression for electrical  conductivity, given by eqn.(4), we have
$\tau_D=10^5-10^6$yrs. Then with the constant $B_m^{(0)}(r)$, we have from eqn.(5)
\begin{equation}
B_m(r,t) =  B_m^{(0)}\left [ 1-\frac{2}{r} (\pi D t)^{1/2}\right ] 
\end{equation}
where we have used $\sigma_n=10^24$sec${-1}$
Hence putting $B_m(r,t)=0$, the estimated time scale for the expulsion 
of magnetic lines is $\sim 10^5-10^6$yrs. Which can also be obtained from stability 
analysis of planer normal-superconducting interface. From this simple analysis, 
we may therefore conclude that, except the time scale for expulsion of magnetic field lines, almost nothing can be inferred, particularly about the 
growth of superconducting zone and the associated expulsion of magnetic flux lines 
from this region by solving eqn.(2). 
The reason behind such uncertainty is our lack of knowledge or definite
ideas on the numerical values of the parameters present in eqn.(6). To get some 
idea of magnetic field expulsion time scale and the structure of the
growing superconducting zone, we shall 
now investigate the morphological instability of normal-superconducting interface 
of  quark  matter using the idea of solute diffusion during solidification of 
alloys. Since the  motion  of normal-superconducting  interface  is extremely 
important in this case and has to be taken into consideration, then instead of 
eqn.(2) which  is valid  in  the  rest frame, an equation expressed in a 
coordinate system which is moving with an element of the boundary  layer  is
the  correct description of such superconducting growth, known as
{\it{Directional Growth}}. The equation is called
{\it{Directional Growth Equation}}, and is given by
\begin{equation}
\frac{\partial  B_m}{\partial  t}  -v  \frac{\partial B_m}
{\partial z} =D\nabla^2 B_m
\end{equation}
where  the motion of the plane interface is assumed to be along z-axis and 
$v$ is the velocity of  the  front.  This  diffusion  equation  must  be
supplemented  by  the  boundary  conditions at the interface. The
first boundary condition is obtained by  combining  Ampere's
and Faraday's laws at the interface, and is given by
\begin{equation}
B_mv\mid_s=-D(\nabla B_m) .\hat n \mid_s 
\end{equation}
where   $\hat   n$  is  the  unit  vector normal to the interface
directed from the normal phase to the superconducting phase. This
is nothing but the continuity equation for magnetic flux diffusion. The  rate
at  which  excess  magnetic  field  lines  are  rejected from the
interior of the phase is balanced by the rate at  which  magnetic
flux lines diffuses ahead of the two-phase interface. This effect
makes  the  boundary  layer  between superconducting-normal quark
matter phases unstable due to excess magnetic field lines present
on the surface  of  the  growing  superconducting  bubble.  Local
thermodynamic  equilibrium at the interface gives (Gibbs-Thompson
criterion)
\begin{equation}
B_m \mid_s  \approx  B_m^{(c)}  \left  (  1-\frac{4\pi   \alpha
}{RB_m^{(c)^2}}\right )
=  B_m^{(c)}  \left  (  1- \delta C \right )
\end{equation}
where  $\delta$  is  called  capillary  length  with $\alpha$ the
surface tension, $C$ is the curvature  $=1/R$  (for  a  spherical
surface), and $B_m^{(c)}$ is the thermodynamic critical field.

To investigate the stability of superconducting-normal interface,
we  shall  follow  the original work by Mullins and Sekerka \cite{R26a,R26b},
and consider a steady state growth of superconducting core,  then
the  time  derivative  in  eqn.(7)  will  not appear. Introducing
$r_\perp=(x^2+y^2)^{1/2}$ as the transverse  coordinate  at  the
interface, then we have after rearranging eqn.(7)
\begin{equation}
\left      [\frac{\partial^2      }{\partial      r_\perp^2}+
\frac{1}{r_\perp}\frac{\partial}{\partial r_\perp}+
\frac{\partial^2 }{\partial z^2}+
\frac{v}   {D}\frac{\partial   }{\partial  z}\right  ]  B_m=0
\end{equation}
The approximation that the solidification is occurring under steady state condition
used in the freezing of molten material will be followed in the present case
of normal to superconducting phase transition. Now if it is
assumed that these two phenomena taking place in 
two completely separate physical world are almost
identical natural processes, then the concentration of magnetic flux 
lines and normal-superconducting interface morphology will be independent of
time. The main disadvantage of this assumption  is  that there will be no
topological evolution  of  the  interface shape. As a consequence of this
constraint the solution  to the basic  diffusion  problem
is   indeterminate   and   a   whole  range  of  morphologies  is
permissible from the mathematical point  of  view.  In  order  to
distinguish  the  solution which is the most likely to correspond
to reality, it is necessary to  find  some  additional  criteria.
The examination  of  the  stability  of a slightly perturbed growth
form is probably the most reasonable manner in which
this situation may be treated. In the following we shall investigate  the
morphological  instability  of  normal-superconducting  interface
from eqn.(10). Assuming a solution of this equation expressed as the product of
separate functions of $r_\perp$ and $z$ and setting the
separation  constant  equal  to  zero and  using the boundary
condition given by eqn.(9), we have for an  unperturbed  boundary
layer moving along $z$-axis
\begin{equation}
B_m=B_m^{(s)} \exp(-zv/D) =B_m^{(s)} \exp(-2z/l)
\end{equation}
where $l=2D/v$ is the layer thickness, which is  very  small  for
small $D$. Mathematically, the thickness of this layer is infinity.
For practical purpose an effective value $l$ can  be  taken.  The
order  of  magnitude  estimates  or limiting values for the three
quantities $D$, $v$ and $l$ can be obtained  from  the  stability
condition of planer interface, which will be discussed latter.

Due to excess magnetic flux lines at the interface, the form of
the planer  normal-superconducting  interface  described  by  the
equation  $z=0$  is assumed to be changed by a small perturbation
represented by the simple sine function
\begin{equation}
z=\epsilon \sin(\vec k.\vec r_\perp)
\end{equation}
where  $\epsilon$  is  very  small  amplitude and $\vec k$ is the wave
vector of the perturbation. Then the perturbed  solution  of  the
magnetic field distribution near the interface can be written as
\begin{equation}
B_m=B_m^{(s)}\exp(-vz/D)      +A\epsilon     \sin(\vec     k.\vec
r_\perp)\exp(-bz)
\end{equation}
where  $A$  and $b$ are two unknown constants. Since the solution
should satisfy the diffusion equation (eqn.(10)), we have
\begin{equation}
b=\frac{v}{2D}+\left     [     \left    (\frac{v}{2D}\right
)^2+k^2\right ]^{1/2}
\end{equation}
To  evaluate  $A$,  we utilise the assumption that $\epsilon$ and
$\epsilon \sin(\vec k.\vec r_\perp)$ are small enough so that  we
can  keep  only the linear terms in the expansion of exponentials
present  in  eqn.(13). Then at the interface, we have after some straight forward 
algebraic manipulation
\begin{equation}
A=\frac{v}{D} B_m^{(s)}
\end{equation}
The  expression  describing the magnetic field distribution ahead
of the slightly perturbed interface then reduces to
\begin{equation}
B_m=B_m^{(s)} \left [ \exp(-vz/D)  +\frac{v}{D}  \epsilon  \sin
(\vec k.\vec r_\perp)\exp(-bz) \right ]
\end{equation}

Now from the other boundary condition (eqn.(9)) we have
\begin{equation}
B_m^{(s)}=B_m^{(c)}-\frac{4\pi \alpha B_m^{(c)}}{B_m^{(s)2}}C
\end{equation}
where $C=z^{''}/(1+z^{'2})^{3/2}$ is the curvature at $z=\epsilon
\sin(\vec  k.\vec  r_\perp)$  and prime indicates derivative with
respect to $r_\perp$.

Neglecting $z^{'2}$, which is small for small perturbation, we have
\begin{equation}
B_m^{(s)}=B_m^{(c)} +\Gamma k^2 S
\end{equation}
where $\Gamma =4\pi  \alpha  B_m^{(c)}/B_m^{(s)2}$  and  we  have
replaced  $\epsilon  \sin(\vec k.\vec r_\perp)$ by $S$. Since the
amplitude of perturbation  $\epsilon$  is  extremely  small,  the
quantity $S$ is also negligibly small.

Now the eqn.(18) is also given by
\begin{equation}
B_m^{(s)}= B_m^{(c)}+ GS
\end{equation}
where
\begin{equation}
G=\frac{dB_m}{dz}\mid_{z=S}     =-\frac{v}{D}    \left    (
1-\frac{vS}{D} \right ) B_m^{(s)} -bAS (1-bS) 
\end{equation}
Combining these two equations, we have
\begin{equation}
k^2\Gamma +\frac{v}{D}  \left  (  1-\frac{vS}{D}  \right  )
B_m^{(s)} -\frac{bv}{D} B_m^{(s)} S(1-bS)=0
\end{equation}
This  expression determines the form  (values of $k$) which the
perturbed interface must assume in order to satisfy  all  of  the
conditions  of the problem. To analyse the behaviour of the roots,
we replace right hand side of eqn.(21) by  some  parameter  $-P$.
(We  have  taken  $-P$  in order to draw a close analogy with the
method given in refs.\cite{R25,R25a}). Then rearranging eqn.(21), we have
\begin{equation}
-k^2\Gamma  -\frac{v}{D}  \left  (  1-\frac{vS}{D} \right )
B_m^{(s)} +\frac{bvB_m^{(s)} S}{D} (1-bS) =P
\end{equation}
(in  refs.\cite{R25,R25a}  the  parameter  $P$  is  related  to  the time
derivative of $\epsilon$, the amplitude of small perturbation). If
the  parameter  $P$  is  positive  for  any  value  of  $k$,  the
distortion  of  the  interface  will  increase,  whereas, if it is
negative for all values of $k$, the perturbation  will  disappear
and  the interface will be stable. In order to derive a stability
criterion, it only needs to know whether eqn.(20) has roots  for
positive values of $k$. If it has no roots, then the interface is
stable  because  the  $P-k$  curve never rises above the positive
$k$-axis and $P$ is therefore negative for  all  wavelengths.  We
have  used Decarte's theorem to check how many positive roots are
there. It is more convenient to express $k$ in terms of  $b$  and
then replacing $b$ by $\omega +v/D$, we have from eqn.(22)
\begin{eqnarray}
-\omega^2 \left ( \Gamma +\frac{vB_m^{(s)} S^2}{D}  \right  )
&-& \omega  \left ( \Gamma +\frac{2vB_m^{(s)} S^2}{D} -B_m^{(s)} S
\right  )  \frac{v}{D}  \nonumber\\ &-&\frac{v}{D}  B_m^{(s)} 
\left  (  1- \frac{v}{D} S\right )^2 =P
\end{eqnarray}
This  is  a  quadratic equation for $\omega$. The first and the
third terms are always negative. The second  term  will  also  be
negative if
\begin{equation}
\Gamma + \frac{2vB_m^{(s)} S^2}{D} -B_m^{(s)} S>0
\end{equation}
Then  it follows from Decart's rule that if the condition (25) is
satisfied, there can not be any positive root. Which implies that
the small perturbation of the interface will disappear. Since the
amplitude of perturbation is assumed to be extremely  small,  the
quantity   $S=\epsilon   \sin(\vec  k  .\vec  r_\perp)$  is  also
negligibly small. Under such circumstances  the  middle  term  of
eqn.(23)  is  much  smaller  than rest of the terms. The Decart's
rule given by the condition (24) can be re-written as
\begin{equation}
\Gamma > B_m^{(s)} S
\end{equation}
Which after some simplification gives the stability criterion for
the plane unperturbed interface, given by
\begin{equation}
\alpha >\frac{B_m^{(s)3}S}{4\pi B_m^{(c)}}
\end{equation}

From   the   stability   criterion,   it   follows    that    the
normal-superconducting  interface energy/area of quark matter has
a lower bound, which depends  on  the  interface  magnetic  field
strength,  critical  field  strength and also on the perturbation
term $S$. An order  of  magnitude  of  this  lower  limit  can  be
obtained by assuming $B_m^{(s)}=10^{-3}B_m^{(c)}$.
(Since  the  critical  field  $B_m^{(c)}\sim  10^{16}$G,  and the
neutron star magnetic field strength $B_m\sim 10^{13}$G,  we  may
use this equality). Then the lower limit is given by
\begin{equation}
\alpha_L\approx  10^{-9}~~{\rm{MeV/fm}}^2~\left( \frac{S}{{\rm
{fm}}}\right )
\end{equation}
On the other hand for $B_m^{(s)}=0.1 B_m^{(c)}$, we have
\begin{equation}
\alpha_L\approx  10^{-3}~~{\rm{MeV/fm}}^2~\left( \frac{S}{{\rm
{fm}}}\right )
\end{equation}
The approximate general expression for the lower limit  is  given
by
\begin{equation}
\alpha_L\approx  h^3~{\rm{MeV/fm}}^2~\left( \frac{S}{{\rm
{fm}}}\right )
\end{equation}
where  $h=B_m^{(s)}/B_m^{(c)}$.  Therefore  the maximum value of
this lower limit is
\begin{equation}
\alpha_L^{\rm{max.}}\approx  1~{\rm{MeV/fm}}^2~\left( \frac{S}{{\rm
{fm}}}\right )
\end{equation}
when the two phase are in thermodynamic  equilibrium.  Of  course
for  such  a strong magnetic field, as we have seen \cite{R6,R7} that 
there can not be a first order quark-hadron phase  transition.

On  the  other  hand  if  we do not have control on the interface
energy, which can in principle be obtained  from  Landau-Ginzberg
model,   we  can  re-write  the  stability  criteria  in terms  of
interface concentration of magnetic field  strength  $B_m^{(s)}$,
and is given by
\begin{equation}
B_m^{(s)}  < \left [ \frac{4\pi \alpha B_m^{(c)} }{S\left ( 1-
\frac{2v}{D}S\right )} \right ]^{1/3}
\end{equation}
This  is more realistic than the condition imposed on the surface
tension $\alpha$. Now for a type-I  superconductor,  the  surface
tension  $\alpha  >0$,  which implies $1-2vS/D >0$. Therefore, we
have  $2vS/D  <  1$,  and  for  the   typical   value   $\sigma_n
\sim10^{26}$ sec$^{-1}$ for the electrical conductivity of normal
quark  matter, the profile velocity $v< D/2S \sim 10^{-6}/S$cm/sec 
$\sim 1$cm/sec for  $S\sim  10^{-6}$cm. Therefore the interface velocity 
$<1$cm/sec for such typical values  of $\sigma_n$ and $S$ to make the planer 
interface stable under small perturbation. Now the thickness of the layer  at  the
interface   is  $l=2D/v>10^{-6}$cm  for  such  values  of  $D$  (or
$\sigma_n$)  and  $v$.  Here  $S$  is  always  greater  than  $0$,
otherwise,     the     magnetic    field    strength    at    the
normal-superconductor interface becomes unphysical. As before, if
the second term of eqn.(23) is negligibly small compared to other
two terms, we have
\begin{equation}
B_m^{(s)}  <  \left  [  \frac{4\pi \alpha B_m^{(c)}}{S}\right
]^{1/3} 
\end{equation}

\section{Conclusion}
Therefore  we  may  conclude  that if a superconducting
transition occurs in a quark star, the magnetic properties
of such bulk object are entirely different from that of  a  small
laboratory  superconducting  sample.  Expulsion  of magnetic flux
lines from the superconducting zone  is  not at all instantaneous.  The
typical  time scale is $10^5-10^6$ yrs. 
We have noticed that this time scale is
very close to the magnetic field decay time scale in a neutron star.
Due to the presence of excess
magnetic flux lines at the interface, which is true  if
the  diffusion  rate  of  magnetic  lines of forces in the normal
phase is less than the rate  of  growth  of  the  superconducting
zone, the topological structure of normal-superconducting boundary layer
may  change significantly. Of course, it depends on the magnitude
of surface tension $\alpha$. It may take dendritic shape  instead
of  planer  structure.  The  stability  of  planer interface also
depends on the strength of interface magnetic field at the boundary layer. 
Since the expulsion time scale is very high, we expect that there will be no 
instability at the interface between normal and superconducting quark matter
phase. The superconducting phase will grow steadily.
How to get evidence from observational data for such an unusual
shape of normal-superconducting quark matter interface 
is a matter of further study.

Finally, we know that in the laboratory, since the size of the
superconducting sample is about a few cm, the expulsion is instantaneous, if
there is a type one transition to superconducting phase. On the other hand,
the problem we are dealing with is having a linear dimension of a few Km,
therefore, it is quite obvious, that the expulsion can not be instantaneous
nature and the model of impurity diffusion in molten alloy, we are employing here
is one of the techniques to study such transition in bulk phase. We believe,
that the model is also applicable to any bulk system of charged fermions,
where a type one super-conducting transition is occurring.


\begin{thebibliography}{99}
\bibitem{R1}
Witten E., 1984, Phys. Rev. D30, 272.
\bibitem{R2}
 Truemper J.,  et.  al., 1978, Ap.  J.  219, L105. 
\bibitem{R3}
 Wheaton W.A., et. al., 1979, Nature 272, 240. 
\bibitem{R4}
Gruber D.E., et. al., 1980, Ap. J.  240, L127. 
\bibitem{R5}
Mihara T., et. al., 1990, Nature 346, 250.
\bibitem{R6}
Chakrabarty S., 1995, Phys. Rev. D51, 4591.
\bibitem{R7}
Ghosh T. and Chakrabarty S., 2001, Phys. Rev. D63 043006.
\bibitem{R8}
Fetter A.L. and  Walecha J.D., 1971, Quantum Theory of Many
Particle System, McGraw Hill Book Company, New York.
\bibitem{R9}
Baym G., Pethick C.  and  Pines D., 1969,  Nature  224,  673.
\bibitem{R10}
Shapiro S.L. and Teukolsky S.A., 1983, Black Holes, White
Dwarfs and Neutron Stars, Wiley, New York.
\bibitem{R11}
Baldo M., et. al., 1992, Nucl. Phys. A536, 349.
\bibitem{R12}
Bailin D. and Love A., 1984, Phys. Rep. 107, 325.
\bibitem{R13}
 Proceedings of the NATO advanced research workshop on
"Superdense QCD matter and compact stars", Yerevan, Armenia, September
27- October 4, 2003. Nato Science Series II:Mathematics,
Physics and Chemistry, Vol. 197, Springer 2006, D. Blaschke and D.
Sedrakian (eds.).
\bibitem{R14}
Alford M., Annual Review of Nuclear and Particle Science, 2001, 51, 131.
\bibitem{R15}
Blaschke D., Voskresensky D.N. and  Grigorian H., 2005, hep-ph/0510368.
\bibitem{R16}
Rajagopal K. and Wilczek F., 2000, hep-ph/0011333.
\bibitem{R17}
Alford M.G., Rajagopal K. and Wilczek F., 1999 Nucl. Phys.  B537, 443.
\bibitem{R18}
Alford M.G, Kouvaris C. and  Rajagopal K., 2004, Phys. Rev.  Lett. 92, 222001.
\bibitem{R19}
Page D, Lattimer J.M., Prakash M. and Steiner A.W., Ap., 2004, J. Suppl. 
155, 623. 
\bibitem{R19a} Grigorian H., Blaschke D. and  Voskresensky D.N., 
2005, Phys. Rev. C71, 045801.
\bibitem{R20}
Konenkov D. and Geppert U., 2000, MNRAS 313, 66;
2000, MNRAS 325, 426.
\bibitem{R22}
Bailin B. and  Love A., 1982, Nucl. Phys. B205, 119.
\bibitem{R23}
Chau A.F., 1997,  Ap. J. 479, 886.
\bibitem{R24}
Alford M.G., Berges J. and  Rajagopal K., 2000,  Nucl. Phys. B571, 269.
\bibitem{R24a} Losert W., Shi B.Q. and Cummins H.Z., 1998, Proc. Natl. Acad.
Sci. USA (Applied Physical Science) 95, 431.
\bibitem{R25}
Adams F.C, Freese K. and  Langer J.S., 1993, Phys. Rev.
D47, 4303.  
\bibitem{R25a}
Kamionkowski Marc and Freese K., 1992, Phys.
Rev. Lett. 69, 2743.
\bibitem{R26a}
Mullins W.W. and  Sekerka R.F., 1963, Jour. Appl. Phys. 34,
323.
\bibitem{R26b}
Mullins W.W. and  Sekerka R.F., 1964, Jour. Appl. Phys.
1964; 35, 444.
\bibitem{R27}
Langer J.S., 1980, Rev. Mod. Phys.  52,  1.
\bibitem{R28}
Haensel P. and  Jerzak A.J., 1989, Acta Phys. Pol. B20, 141.
\bibitem{R28a}
Haensel P., 1991, Nucl. Phys. B24 (proc. of the  Int.  Workshop
on  Strange  quark  Matter  in Physics and Astrophysics, Univ. of
Aarhus, Denmark, May 20-24, 1991), 23.
\bibitem{D48}  Heiselberg H and Pethick C.J, 1993, Phys. Rev. D48, 2916.
\bibitem{R29}
Chakrabarty S., Bandopadhyay D. and  Pal S, 1997, Phys. Rev.
Lett. 78, 2898.
\bibitem{R29a}
Bandopadhyay D, Chakrabarty S. and  Pal S., 1997, Phys. Rev.
Lett. 79, 2176.
\end{thebibliography}
\end{document}